\documentclass[a4paper, reqno]{amsart}
\usepackage{graphicx}
\usepackage{natbib}
\usepackage{authblk}
\usepackage{rotating}

\usepackage{amsmath}
\usepackage{amssymb}
\usepackage{amsthm}
\usepackage{amsaddr}

\numberwithin{equation}{section}

\calclayout

\begin{document}
\title[Mechanistic model of disjunctive metabolic symbioses]{A mechanistic model of disjunctive metabolic symbioses in microbes}
\author{Matthias M. Fischer}
\address{Freie Universit\"at Berlin; Fachbereich Biologie, Chemie, Pharmazie; Institut f\"ur Biologie, Mikrobiologie; 14195 Berlin, Germany. Tel.: +49 30 838 53373}
\email{m.m.fischer@fu-berlin.de}
\date{\today}

\maketitle

\begin{abstract}
Lately, experimental research on microbial symbioses based on nutrient exchange and interdependence has yielded a number of interesting findings, however an in-depth mathematical description of the exact underlying dynamics of such symbiotic associations is still missing. Here, we derive and analyse a mechanistic mathematical model of such a relationship in a continuous chemostat culture based on five coupled differential equations. The influence of the biological traits of the involved organisms on the position and stability of the equilibrium states of the system is examined. We also demonstrate how manipulating the external metabolite concentrations of the system can shift the species interaction on a continuous spectrum ranging from mutualism over commensalism to parasitism.
\end{abstract}

\section{Introduction}
A symbiosis is defined as a close and usually long-term interaction between two different biological species \citep{def_symbioses}. Generally, three different broad categories of symbiotic relationships have been distinguished: mutualism, in which both individuals derive a benefit; commensalism, where one individual benefits and the other one does experience neither harm nor benefit; and parasitism, in which one organism, termed the parasite, benefits at the expense of its symbiotic partner, called the host \citep{model_carrying_capacity}. One specific form of symbiosis is called 'syntrophy', a term often used in the field of microbiology to describe a symbiotic relationship between two species of microorganisms based on nutritional interdependence \citep{prescott}. In other words, in such a symbiotic cross-feeding relationship the two species live of and depend on products released into their shared environment by the respective other symbiont. \\

Experimental research in the field of syntrophic mutualism of microbial populations has already yielded a number of interesting results. \textit{Mee} and colleagues have constructed communities of different auxotrophic strains of \textit{Escherichia coli} that relied on amino acids produced by their respective symbiotic partner in order to grow \citep{ecoli_strains_aa_syntrophy}. The authors analysed their experimental results by infering a matrix of pairwise 'cooperativity coefficients' that quantify how the presence of one specific auxotrophic strain influenced the growth of the respective other auxotrophic strains, but did not characterise the population sizes and amino acid concentrations as functions of time any further. \textit{Kerner} and colleagues also have implemented such a 'forced symbiosis' using the same approach with a pair of \textit{E. coli} strains which were auxotroph for the amino acids tryptophan and tyrosine respectively \citep{tunable}. They analysed the behaviour of their symbiotic system by devising a set of four coupled differential equations for the two population densities and the two amino acid concentrations over time. The authors were able to derive an analytical formula describing the final equilibrium state of the system as functions of the amino acid efflux rates and demands of the two symbionts. However, this model has been derived under the assumption of an identical behaviour of the two strains with regard to their resource consumption, resource affinity and intrinsic growth rates, and was not generalised to cases in which the two symbionts differ with regard to these parameters. \\

A study of importance for this work was carried out by \textit{Megee} and colleagues, who experimentally studied a 'symbiotic spectrum' for a two-species syntrophic culture consisting of the baker's yeast \textit{Saccharornyces cerevisiae} and a strain of the bacterium \textit{Lactobacillus casei}. Both microbes competed for the limiting resource glucose, while the latter was dependant on riboflavin produced by the yeast utilising glucose as carbon source. The authors noted that \textit{the importance of the abiotic environment in such studies can hardly be overemphasized} and that \textit{care has been taken to control and (or) measure the more important parameters of the abiotic environment} -- and indeed through careful manipulation of the concentrations of glucose and riboflavin in the growth medium the interaction between the two microbes could be shifted on a spectrum ranging from commensalism over pure competition to mutualism, as demonstrated by measuring growth behaviour and rigorous mathematical analyses. It has to be noted, however, that their mathematical framework was directly aimed at the special case which has been experimentally studied in their work alone, and not at the general case of arbitrary species exchanging arbitrary metabolites \citep{syntrophy_spectrum}. A more recent work has studied this topic both experimentally by cocultivating two free-floating metabolically interdependent yeast strains which exchange amino acids and by using strongly simplified and rather abstract 'phenomenological models' based on two coupled ordinary differential equations. The authors came to a similar conclusion and demonstrated how varying external concentrations of the exchanged amino acids can qualitatively change the nature of the species interaction among a 'symbiotic spectrum' ranging from obligate mutualism to competition and competitive exclusion as the extreme ends \citep{yeast_strains_metabolite_concentrations}.\\

Purely theoretical reflections on such symbiotic relationships have lead to a number of different mathematical approaches such as the model by \textit{Yukalov} and colleagues that describes symbioses as a mutual influence of species on the livelihoods and thereby the carrying capacities of each other \citep{model_carrying_capacity}. Another mathematical model by \textit{Graves and Peckham} understands symbiotic relationships as species mutually influencing their realised growth rates in a given environment \citep{model_growth_rates}. While simple and mathematically elegant, these models are inherently phenomenological in nature, which means that they reproduce observed behaviour of a study system without providing any further mechanistic insight into the underlying processes of the symbiotic relationship. Moreover, these models contain a number of parameters such as so-called 'symbiosis coefficients' that need to be chosen rather arbitrarily and cannot be directly understood from first principles, but only inferred from data \textit{a posteriori}. This makes understanding them from a biological point of view difficult. \\

Hence, in this study a mechanistic model of the syntrophic relationship between two microbial species was developed and analysed containing only parameters which can be measured experimentally and directly manipulated. The rest of the paper is organised as follows: In Section \ref{sec:model} the mathematical model is derived, and in Section \ref{sec:params} we describe its standard parametrisation. In Section \ref{sec:stst} we find the steady states of the system and analyse their stability, afterwards in Section \ref{sec:conditions} we study the effect of differences in the biological properties on position and stability of the equilibria of the system. Additionally, in Section \ref{sec:spec} we show how manipulating the external metabolite concentrations in the medium can shift the species interaction on a continuous spectrum ranging from mutualism over commensalism to parasitism. Finally, in Section \ref{sec:disc} we discuss some biological implications of the results of this paper.

\section{Theoretical model}
\label{sec:model}
The devised model relies on the following assumptions:

\begin{enumerate}
\item{The population densities $N_1(t)$ and $N_2(t)$ of the two interacting species depend on respectively one metabolite emitted by the respective other species following a \textit{Monod} growth kinetic \citep{monod}. Let $K_1, K_2$ denote the associated \textit{Monod} constants of the two species. The concentrations of the two metabolites in the growth medium are denoted $M_1(t)$ and $M_2(t)$.}
\item{One individual of $N_1$ ($N_2$) changes the concentration of the metabolite $M_1$ ($M_2$) at a rate of $\epsilon_1$ ($\epsilon_2$) due to metabolite efflux into the culture medium.}
\item{Additionally, the growth of both populations relies on a limited resource with concentration $R(t)$, following \textit{Monod} kinetics, as well. Let $L_1, L_2$ denote the associated \textit{Monod} constants of the two species.}
\item{Each newly arising individual of $N_1$ ($N_2$) reduces the metabolite concentration by $\gamma_1$ ($\gamma_2$) and the concentration of the limited resource by $\alpha_1$ ($\alpha_2$). }
\item{Substances and cells are washed out (or diluted) with a fixed rate $\omega$ and the culture is supplied with the limited resource at the same rate, where $R_{in}$ denotes the influx concentration. If desired, metabolites can be added into the culture as well, if so, their influx concentrations are denoted as $M_{1,in}, M_{2,in}$.}
\end{enumerate}

\begin{figure}[h!]
	\centering
	\includegraphics[width=8cm]{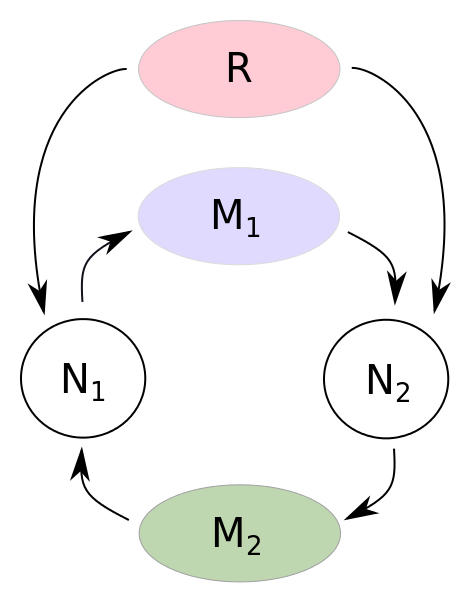}
	\caption{\textit{Schematic overview over the devised model describing the disjunctive symbiosis based on metabolic interdependence. Species $N_1$ releases metabolite $M_1$ into the culture medium which is taken up by species $N_2$, which in turn releases metabolite $M_2$ into the medium which is required by species $N_1$. Both $N_1$ and $N_2$ depend on a limited resource $R$ as well.}}
	\label{fig:ode1}
\end{figure}

Figure \ref{fig:ode1} provides a schematic overview of the modelled system. \\

The final model consists of a system of five coupled ordinary differential equations. The effective growth rates $\beta_1$ and $\beta_2$ of the two populations are functions of resource and metabolite availability and amount to

\begin{equation}
\begin{split}
\label{ode1_betas}
\beta_1 (M_2(t), R(t)) &= \frac{M_2(t)}{M_2(t) + K_1} \frac{R(t)}{R(t)+L_1} r_1, \\
\beta_2 (M_1(t), R(t)) &= \frac{M_1(t)}{M_1(t) + K_2} \frac{R(t)}{R(t)+L_2} r_2.
\end{split}
\end{equation}

Accordingly, the growth of the two population densities are given by

\begin{equation}
\label{ode1_popsizes}
\begin{split}
\dot N_1(t) &= (\beta_1(M_2(t), R(t)) - \omega) \cdot N_1(t) \\
\dot N_2(t) &= (\beta_2(M_1(t), R(t)) - \omega) \cdot N_2(t). 
\end{split}
\end{equation}

The change in resource concentration is therefore given by

\begin{equation}
\label{ode1_R}
\begin{split}
\dot R(t) &= \omega \cdot (R_{in} - R(t)) - \alpha_1 \cdot  \beta_1(M_2(t), R(t)) \cdot N_1(t) - \alpha_2 \cdot  \beta_2(M_1(t), R(t)) \cdot N_2(t).
\end{split}
\end{equation}

Finally, the changes in metabolite concentrations amount to
\begin{equation}
\label{ode1_M12}
\begin{split}
\dot M_1(t) &=  \omega \cdot (M_{1, in} -
M_1(t)) + \epsilon_1 N_1(t) - \gamma_2 \cdot  \beta_2(M_1(t), R(t)) \cdot  N_2(t),  \\
\dot M_2(t) &= \omega \cdot (M_{2, in} - M_2(t)) + \epsilon_2 N_2(t) - \gamma_1 \cdot \beta_1(M_2(t), R(t)) \cdot N_1(t). \\
\end{split}
\end{equation}

\section{Parametrisation of the model}
\label{sec:params}
The model was parametrised as described in Table \ref{tab:params} unless stated otherwise. If the limited resource is thought to be a carbon source such as glucose, an influx concentration of 2 $g/l$ is a realistic quantity which is often encountered in various nutrient broth formulations such as M9 minimal medium \citep{m9}. The washout rate of the culture is set to 0.1 $h^{-1}$, a typical value within the standard parameter range for long-term chemostat experiments \citep{omega}. \\

In the next section of the table, the biological parameters of the two species are listed. The intrinsic growth rate is set to 1 $h^{-1}$, which equals one cellular division per hour, a value typical for bacteria such as \textit{E. coli} in minimal nutrient medium \citep{brockmadigan}. Due to the lack of research literature on the topic, the efflux rate had to be estimated to be 300 $fg/h$. The yield factor of the cells for the exchanged metabolites and the associated \textit{Monod} constants were assumed to be similar to the ones of \textit{E. coli} for the nutrient phosphate ($PO_4^{3-}$). Therefore, a yield factor $\gamma$ of 100 $fg$ per cell \citep{ecoli_composition} and a \textit{Monod} constant of 10 $\mu g$ per single cell \citep{ecoli_phosphate_monod} is used. \\

 The amount $\alpha$ of the limited resource $R$ required to build one new cell is estimated to be approximately 1000 $fg$: The dry weight of one cell of \textit{E. coli} falls in the range of 100 to 1000 $fg$ per cell depending on the speed of growth \citep{ecoli_dryweight} and the conversion rate from sugar to dry biomass is approximately 0.5 in case of aerobic bacterial growth \citep{ecoli_conversionrate}. Therefore, on average an amount of 1000 $fg$ resource per one new cell is a reasonable estimate. The \textit{Monod} constant $L$ of the cells for the limited resource $R$ is set to 100 $\mu g/ml$, approximately equalling the \textit{Monod} constant of \textit{E. coli} for glucose \citep{ecoli_monod_for_glucose}. \\

\begin{table*}[!h]
\begin{tabular} {l l l l}
\hline \\[-0.7em]
    Model Parameter & Symbol & Value & Unit \\
\hline \\[-0.6em]
	\textbf{Culture conditions} &&& \\
	Influx concentration limited resource & $R_{in}$ & 2 & \textit{g/l} \\
	Influx concentration metabolite \#1 & $M_{1,in}$ & 0 &  \textit{g/l}  \\
	Influx concentration metabolite \#2 & $M_{2,in}$ & 0 &  \textit{g/l}  \\
	Washout rate & $\omega$ & 0.1 & $h^{-1}$ \\
	
	&&& \\
		
	\textbf{Parameters symbionts} & & & \\
	Intrinsic growth rate & $r_1, r_2$ & 1 & $h^{-1}$ \\[0.5em]
	
	Efflux rate for metabolite \#1, \#2 & $\epsilon_1, \epsilon_2$ & $300$ & $fg/(cell*h)$ \\
	Yield factor for metabolite \#2, \#1 & $\gamma_1, \gamma_2$ & $100$ & \textit{fg/cell} \\
	Monod constant for metabolite \#2, \#1 & $K_1, K_2$ & 10 & \textit{$\mu$g/ml} \\[0.5em]
	
	Yield factor for limited resource & $\alpha_1, \alpha_2$ & $1000$ & \textit{fg/cell} \\
	Monod constant for limited resource & $L_1, L_2$ & 100 & \textit{$\mu$g/ml} \\
\hline  
\end{tabular} \\[1.3em]
\caption{\textit{The standard parametrisation of the model.}}
\label{tab:params}
\end{table*}

\section{Steady states and stability}
\label{sec:stst}
Assume the existence of a non-trivial steady state with $N_1(t) > 0$ and $N_2(t) > 0$ the system converges to with $t \rightarrow \infty$. Then, by definition, we get

\begin{equation}
\dot N_1(\infty) = \dot N_2(\infty) = \dot R(\infty) = \dot M_1(\infty) = \dot M_2(\infty) = 0.
\end{equation}

By letting $\beta_1 = \beta_2 = \omega$, we obtain from equation (\ref{ode1_popsizes})

\begin{equation}
\frac{M_2(\infty)}{M_2(\infty)+K_1} \frac{R(\infty)}{R(\infty)+L_1} r_1 = \frac{M_1(\infty)}{M_1(\infty)+K_2} \frac{R(\infty)}{R(\infty)+L_2} r_2,
\end{equation}

which shows that at the steady state, regardless of the amount of additionally supplemented metabolites, the metabolite concentrations in the medium will be exactly equal, as long as the two symbionts share the same metabolite and resource affinities and intrinsic growth rates. Using the above equation, we can also calculate the ratio of the two metabolite concentrations at the steady state in case of symbionts with different traits. \\

We apply $\beta_1 = \beta_2 = \omega$ to equation (\ref{ode1_R}) and obtain

\begin{equation}
R(\infty) = R_{in} - \alpha_1 N_1(\infty) - \alpha_2 N_2(\infty).
\end{equation}

Using this equation, from $\beta_1 = \beta_2 = \omega$ we also obtain the two following relationships between metabolite concentrations and population densities at the steady state

\begin{equation}
\begin{split}
M_1(\infty) &= K_2 \left(\frac{r_2}{\omega} \frac{R_{in} - \alpha_1 N_1(\infty) - \alpha_2 N_2(\infty)}{R_{in} - \alpha_1 N_1(\infty) - \alpha_2 N_2(\infty) + L_2} - 1 \right) ^{-1}, \\
M_2(\infty) &= K_1 \left(\frac{r_1}{\omega} \frac{R_{in} - \alpha_1 N_1(\infty) - \alpha_2 N_2(\infty)}{R_{in} - \alpha_1 N_1(\infty) - \alpha_2 N_2(\infty) + L_1} - 1 \right) ^{-1}. \\
\end{split}
\end{equation}

Now consider equation (\ref{ode1_M12}) and solve for the metabolite concentrations. We obtain:

\begin{equation}
\begin{split}
\label{eq*}
M_1(\infty) &= M_{1, in} + \frac{\epsilon_1}{\omega} N_1(\infty) - \gamma_2 N_2(\infty), \\
M_2(\infty) &= M_{2, in} + \frac{\epsilon_2}{\omega} N_2(\infty) - \gamma_1 N_1(\infty).
\end{split}
\end{equation}

Equating the two previous sets of equations yields a system of two equations, which can be solved for $N_1(\infty), N_2(\infty)$. Because the general expressions were too complex for a meaningful biological interpretation, we here exemplarily calculate a contrived example, in which the second symbiont has a intrinsic growth rate twice as high as the first one. This yields four possible solutions, however the last one is biologically not possible due to a theoretically negative metabolite concentration $M_2$. To judge their stability, we evaluate the eigenvalues of the Jacobian of the system, which is given by (we here omit all '$(\infty)$'s for brevity purposes):

\clearpage

\begin{sidewaysfigure}
\begin{equation}
.\\[-15em]
\tiny
\textbf{J} = \left[ \begin{matrix}
\dfrac{M_2 r_1 R}{(K_1+M_2)(L_1+R)} - \omega & 0 & \dfrac{L_1 M_2 N_1 r_1}{(K_1+M_2)(L_1+R)^2} & 0 & \dfrac{K_1 N_1 r_1 R}{(K_1+M_2)^2 (L_1+R)} \\[10ex]

0 & \dfrac{M_1 r_2 R}{(K_2+M_1)(L_2+R)} - \omega & \dfrac{L_2 M_1 N_2 r_2}{(K_2+M_1)(L_2+R)^2} & \dfrac{K_2 N_2 r_2 R}{(K_2+M_1)^2 (L_2+R)} & 0 \\[10ex]

- \dfrac{\alpha_1 M_2 R r_1}{(K_1 + M_2)(L_1+R)} & - \dfrac{\alpha_2 M_1 R r_2}{(K_2 + M_1)(L_2+R)} & - \dfrac{\alpha_1 L_1 M_2 N_1 r_1}{(K_1+M_2)(L_1+R)^2} - \dfrac{\alpha_2 L_2 M_1 N_2 r_2}{(K_2+M_1)(L_2+R)^2} - \omega & - \dfrac{\alpha_1 K_2 N_2 R r_2}{(K_2+M_1)^2 (L_2+R)} & - \dfrac{\alpha_2 K_1 N_1 R r_1}{(K_1+M_2)^2 (L_1+R)} \\[10ex]

\epsilon_1 & - \dfrac{\gamma_2 M_1 r_2 R}{(K_2 + M_1)(L_2+R)} & - \dfrac{\gamma_2 L_2 M_1 N_2 r_2}{(K_2+M_1)(L_2+R)^2} & - \dfrac{\gamma_2 K_2 N_2 r_2 R}{(K_2+M_1)^2 (L_2+R)} - \omega & 0 \\[10ex]

- \dfrac{\gamma_1 M_2 r_1 R}{(K_1 + M_2)(L_1+R)} & \epsilon_2 & - \dfrac{\gamma_1 L_1 M_2 N_1 r_1}{(K_1+M_2)(L_1+R)^2} & 0 & - \dfrac{\gamma_1 K_1 N_1 R r_1}{(K_1+M_2)^2 (L_1+R)} - \omega

\end{matrix}
\right]
\end{equation}
\end{sidewaysfigure}

\clearpage
\pagebreak

We obtain two stable solutions, shown in Table \ref{tab:sols2}: the trivial steady state of extinction, and a stable state of coexistence. As one might have expected, in the latter case, the population density of the faster-growing symbiont significantly exceeds the population density of its symbiotic partner, in this case by more than one order of magnitude. Therefore, the system seems to be able to compensate for differences in the biological traits of the two symbionts and stay stable despite such differences. In the next section, the effects of such trait differences on the equilibrium states and stability of the system will be analysed systematically. \\

\begin{table*}[!h]
	\tiny
	\begin{tabular} {l l l l l l l}
		\hline \\[-0.7em]
		$N_1(\infty)$ & $N_2(\infty)$ & $R(\infty)$ & $M_1(\infty)$ & $M_2(\infty)$  & Eigenvalues & Stability \\
		\hline \\[-0.6em]
		$0$ & $0$ & $2 \times 10^{12}$ & $0$ & $0$ & ($-0.1$,$-0.1$,$-0.1$,$-0.1$,$-0.1$) & Stable \\
		$\approx 6.8 \times 10^7$ & $\approx 1.9 \times 10^9$ & $\approx 5.6 \times 10^{9}$ & $\approx 1.0 \times 10^{10}$ & $\approx 5.8 \times 10^{12}$ & ($\approx -0.06 \pm 0.31i$, $\approx -0.1$, $\approx -0.1$, $\approx -33.2$) & Stable \\
		$\approx 1.9 \times 10^5$ & $\approx 3.9 \times 10^5$ & $\approx 2.0 \times 10^{12}$ & $\approx 5.4 \times 10^8$ & $\approx 1.1 \times 10^9$ & ($\approx 0.06$, $\approx -0.05 \pm 0.09i$,  $\approx -0.1 $, $\approx -0.16$) & Unstable \\
		$\approx 1.9 \times 10^9$ & $\approx 5.8 \times 10^7 $ & $\approx 2.6 \times 10^{9}$ & $\approx 5.8 \times 10^{12}$ & $\approx -2.0 \times 10^{10}$ & Not calculated & Not analysed \\		
		\hline  
	\end{tabular} \\[1.3em]
	\caption{\textit{The steady states of the 'asymmetric' example.}}
	\label{tab:sols2}
\end{table*}

Additionally, we observe that, as long as no metabolites are manually added to the culture, the Jacobian of the extinction state $(N_1(\infty)=0, N_2(\infty)=0, R(\infty)=R_{in}, M_1(\infty)=0, M_2(\infty) = 0)$ always reduces to: \\

\begin{equation}
\textbf{J} = \left[ \begin{matrix}
- \omega & 0 & 0 & 0 & 0 \\[1ex]
0 & - \omega & 0 & 0 & 0 \\[1ex]
0 & 0 & - \omega & 0 & 0 \\[1ex]
\epsilon_1 & 0 & 0 & - \omega & 0 \\[1ex]
0 & \epsilon_2 & 0 & 0 & - \omega \\[1ex]
\end{matrix}
\right],
\end{equation}

which has the five eigenvalues $\lambda_{1,2,3,4,5} = -\omega$, and is, therefore, always a stable steady state. \\

If, however, sufficient amount metabolites are manually added to the medium at the extinction state, the occurring extinction equilibrium state $(N_1(\infty)=0, N_2(\infty)=0, R(\infty)=R_{in}, M_1(\infty)=M_{1, in}, M_2(\infty) = M_{2, in})$ will become unstable instead. The Jacobian of the system in this case becomes a lower diagonal matrix with the eigenvalues $\lambda_{1,2,3} = - \omega$ and \\

\begin{equation}
\begin{split}
\lambda_4 &= \dfrac{M_{2, in}}{K_1+M_{2,in}} \dfrac{R_{in}}{L_1 + R_{in}} r_1 - \omega \\
\lambda_5 &= \dfrac{M_{1, in}}{K_2+M_{1,in}} \dfrac{R_{in}}{L_2 + R_{in}} r_2 - \omega.
\end{split}
\end{equation}

Biologically, this means that sufficient amounts of metabolite influx will let the system converge to the stable state state of non-extinction. This can be explained by the fact that increasing the external metabolite concentrations will increase the realised growth rates of the two populations and, therefore, allow them to grow faster than the washout rate $\omega$.

\section{Conditions for a stable equilibrium}
\label{sec:conditions}
If different species interact in nature, it can be expected that they differ in their biological traits such as their intrinsic growth rate or their metabolite efflux, affinity and requirements. In this section, we will systematically examine the influence of such differences on the occurring equilibria and the stability of the system. \\

\subsection{Intrinsic growth rate}
First we study the effect of different intrinsic growth rates of the symbionts, by fixing $r_1$ at the standard value of 1 $h^{-1}$ and computing the non-trivial stable equilibrium of the system for different values of $r_2$. Table \ref{tab:r2} summarises the results. The system has a stable non-extinction equilibrium for values of $r_2$ between approximately 0.11 to 100 $h^{-1}$, higher values were not tested. Below this range, only the extinction state is stable, which is plausible since the dilution rate of the system is fixed at $\omega = 0.1 \; h^{-1}$, so that an intrinsic growth rate above this level is required for establishing a stable population. Generally, higher intrinsic growth rates will lead to higher population densities at the expense of the symbiotic partner, however in case of very small intrinsic growth rates ($r_2 \lessapprox 0.2$), increasing growth rates will also benefit the symbiotic partner, which can be explained by increasing levels of metabolite availability. \\

\begin{figure}[!h]
	\centering
	\includegraphics[width=10cm]{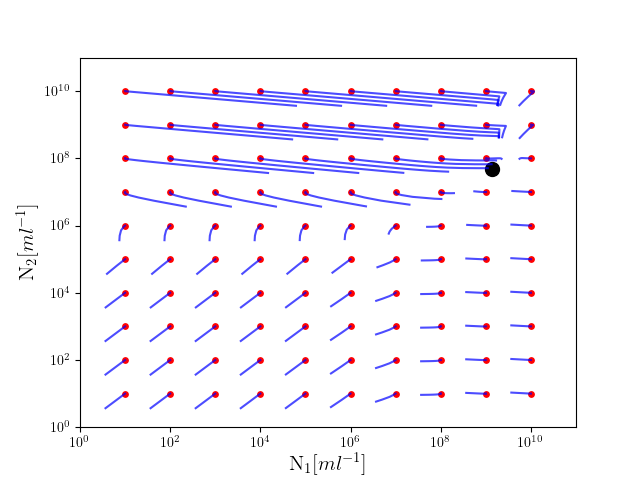}
	\caption{\textit{Phase space diagram of the case of $r_2 = 0.11$. Note, how sufficiently high inoculation sizes are required for reaching the non-extinction state (black disc).}}
	\label{fig:phase_r011}
\end{figure}

As shown previously, the extinction state was found to be stable as well, which suggests that the initial conditions determine whether the system will converge to the extinction or the non-extinction state. We show this with a numerical examination of the case of an extremely low intrinsic growth rate of $r_2 = 0.11$. The phase space of the system with regards to the two population densities is depicted in Figure \ref{fig:phase_r011}. Note how sufficiently high inoculation sizes are required for a stable symbiosis. \\

\begin{table*}[!h]
	\begin{tabular} {l l l l l l}
		\hline \\[-0.7em]
		$r_2$ & $N_1(\infty)$ & $N_2(\infty)$ & $R(\infty)$ & $M_1(\infty)$ & $M_2(\infty)$\\
		\hline \\[-0.6em]
		0.1 & \multicolumn{5}{l}{No non-trivial stable steady state.} \\
		0.11 & $1.4 \times 10^9$ & $4.8 \times 10^7$ & $5.1 \times 10^{11}$ & $4.3 \times 10^{12}$ & $1.2 \times 10^9$ \\
		0.15 & $1.8 \times 10^9$ & $6.2 \times 10^7$ & $1.0 \times 10^{11}$ & $5.5 \times 10^{12}$ & $1.8 \times 10^9$ \\
		0.2 & $1.9 \times 10^9$ & $6.4 \times 10^7$ & $5.0 \times 10^{10}$ & $5.7 \times 10^{12}$ & $2.5 \times 10^9$ \\
		0.25 & $1.9 \times 10^9$ & $6.5 \times 10^7$ & $3.3 \times 10^{10}$ & $5.7 \times 10^{12}$ & $3.3 \times 10^9$\\
		0.5 & $ 1.9 \times 10^9$ & $ 6.7 \times 10^7$ & $ 1.3 \times 10^{10}$ & $ 5.8 \times 10^{12}$ & $ 1.0 \times 10^{10}$ \\
		1 & $ 1.0 \times 10^9$ & $ 1.0 \times 10^9$ & $ 5.6 \times 10^{9}$ & $ 2.9 \times 10^{12}$ & $ 2.9 \times 10^{12}$ \\
		2 & $ 6.8 \times 10^7$ & $ 1.9 \times 10^9$ & $ 5.6 \times 10^{9}$ & $ 1.0 \times 10^{10}$ & $ 5.8 \times 10^{12}$ \\
		5 & $ 6.5 \times 10^7$ & $ 1.9 \times 10^9$ & $ 5.6 \times 10^9$ & $ 2.5 \times 10^{9}$ & $ 5.8 \times 10^{12}$ \\
		10 & $ 6.5 \times 10^7$ & $ 1.9 \times 10^9$ & $ 5.6 \times 10^9$ & $ 1.1 \times 10^{9}$ & $ 5.8 \times 10^{12}$ \\
		100 & $ 6.4 \times 10^7$ & $ 1.9 \times 10^9$ & $ 5.6 \times 10^9$ & $ 1.0 \times 10^{9}$ & $ 5.8 \times 10^{12}$ \\
		\hline  
	\end{tabular} \\[1.3em]
	\caption{\textit{The biologically feasible ($M_1(\infty), M2(\infty) > 0$) non-trivial stable steady state for varying growth rates $r_2$. All values are rounded to the first decimal place.}}
	\label{tab:r2}
\end{table*}

\clearpage

\subsection{Parameters relating to the metabolites}
Now, we analyse how differences in those biological traits, that relate to the efflux and utilisation of the exchanged metabolites, will affect the equilibrium states of the system. \\ 

\begin{figure}[h!]
	\centering
	\includegraphics[width=10cm]{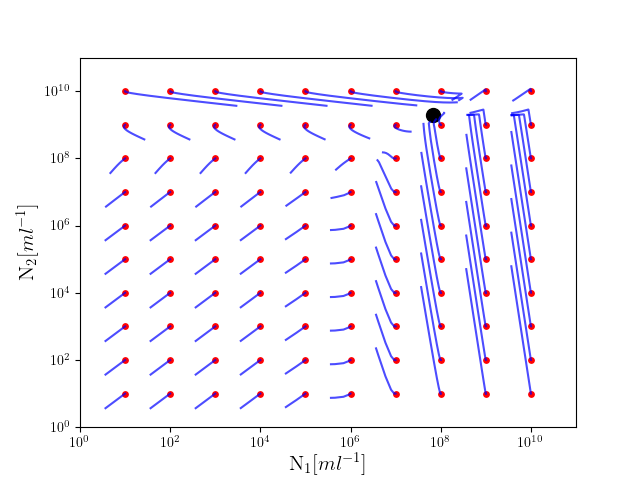}
	\caption{\textit{Phase space diagram of the case of $\epsilon_2 = 1.0$. Note, how sufficiently high inoculation sizes are required for reaching the non-extinction state (black disc).}}
	\label{fig:phase_eps1}
\end{figure}

First, we estimate the range of metabolite efflux rates $\epsilon_2$, in which a stable symbiosis can occur (Table \ref{tab:epsilon 2}). Interestingly, even minor metabolite efflux rates are able to sustain a stable symbiosis, however with significantly asymmetric population densities. Increasing the metabolite efflux rate of a species shifts the equilibrium towards the respective other symbiotic partner. We again examine the system numerically, here the case of an extremely low metabolite efflux of $\epsilon_2 = 1.0$, in order to see how the initial population sizes determine, whether the stable steady state of symbiosis, or the stable extinction state will be reached. Again, we observe that sufficiently high inoculation sizes are required for reaching the former (Figure \ref{fig:phase_eps1}). \\

\begin{table*}[!h]
	\begin{tabular} {l l l l l l}
		\hline \\[-0.7em]
		$\epsilon_2$ & $N_1(\infty)$ & $N_2(\infty)$ & $R(\infty)$ & $M_1(\infty)$ & $M_2(\infty)$\\
		\hline \\[-0.6em]
		0 & \multicolumn{5}{l}{No non-trivial stable steady state.} \\
		1 & $6.8 \times 10^7$ & $1.9 \times 10^9$ & $1.1 \times 10^{10}$ & $1.2 \times 10^{10}$ & $1.2 \times 10^{10}$ \\
		$1 \times 10^1$ & $1.2 \times 10^8$ & $1.9 \times 10^9$ & $5.9 \times 10^{9}$ & $1.8 \times 10^{11}$ & $1.8 \times 10^{11}$ \\
		$1 \times 10^2$ & $5.2 \times 10^8$ & $1.5 \times 10^9$ & $5.6 \times 10^{9}$ & $1.4 \times 10^{12}$ & $1.4 \times 10^{12}$ \\
		$1 \times 10^3$ & $1.5 \times 10^9$ & $ 4.7 \times 10^8$ & $ 5.6 \times 10^{9}$ & $ 4.5 \times 10^{12}$ & $ 4.5 \times 10^{12}$ \\
		$1 \times 10^4$ & $1.9 \times 10^9$ & $ 6.0 \times 10^7$ & $ 5.6 \times 10^{9}$ & $ 5.8 \times 10^{12}$ & $ 5.8 \times 10^{12}$ \\
		\hline  
		
	\end{tabular} \\[1.3em]
	\caption{\textit{The biologically feasible ($M_1(\infty), M2(\infty) > 0$) non-trivial stable steady state for varying metabolite efflux rates $\epsilon_2$. All values are rounded to the first decimal place.}}
	\label{tab:epsilon 2}
\end{table*}

\begin{figure}[h!]
	\centering
	\includegraphics[width=10cm]{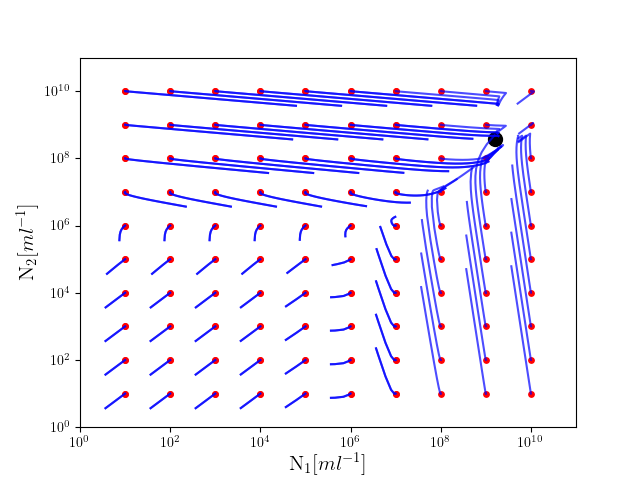}
	\caption{\textit{Phase space diagram of the case of $\gamma_2 = 1 \times 10^4$. Note, how sufficiently high inoculation sizes are required for reaching the non-extinction state (black disc).}}
	\label{fig:phase_gamma10000}
\end{figure}

Next, we use the same method in order to examine the influence of differences in metabolite requirements of the two symbionts on the equilibria and stability of the system (Table \ref{tab:gamma 2}). Increasing metabolite requirements shift the equilibrium states towards lower population sizes of the respective species, whereas the population density of the other symbiont grows. We once again numerically compute the phase space diagram of the most extreme case analysed ($\gamma_2 = 1 \times 10^4$) to demonstrate how the initial population sizes determine which stable steady state will be reached (Figure \ref{fig:phase_gamma10000}). \\

\begin{table*}[!h]
	\begin{tabular} {l l l l l l}
		\hline \\[-0.7em]
		$\gamma_2$ & $N_1(\infty)$ & $N_2(\infty)$ & $R(\infty)$ & $M_1(\infty)$ & $M_2(\infty)$ \\
		\hline \\[-0.6em]
		1 & $9.8 \times 10^8$ & $ 1.0 \times 10^9$ & $ 5.6 \times 10^{9}$ & $ 2.9 \times 10^{12}$ & $2.9 \times 10^{12}$ \\
		10 & $ 9.8 \times 10^8$ & $ 1.0 \times 10^9$ & $ 5.6 \times 10^{9}$ & $ 2.9 \times 10^{12}$ & $2.9 \times 10^{12}$ \\
		100 & $ 1.0 \times 10^9$ & $ 1.0 \times 10^9$ & $ 5.6 \times 10^{9}$ & $ 2.9 \times 10^{12}$ & $2.9 \times 10^{12}$ \\
		1000 & $1.1 \times 10^9$ & $8.7 \times 10^8$ & $5.6 \times 10^9$ & $2.5 \times 10^{12}$ & $2.5 \times 10^{12}$ \\
		10000 & $1.6 \times 10^9$ & $3.8 \times 10^8$ & $5.6 \times 10^9$ & $9.9 \times 10^{11}$ & $9.9 \times 10^{11}$ \\	
		\hline  		
	\end{tabular} \\[1.3em]
	\caption{\textit{The biologically feasible ($M_1(\infty), M2(\infty) > 0$) non-trivial stable steady state for varying metabolite requirements $\gamma_2$ per new cell of the second species. All values are rounded to the first decimal place.}}
	\label{tab:gamma 2}
\end{table*}

\begin{figure}[h!]
	\centering
	\includegraphics[width=10cm]{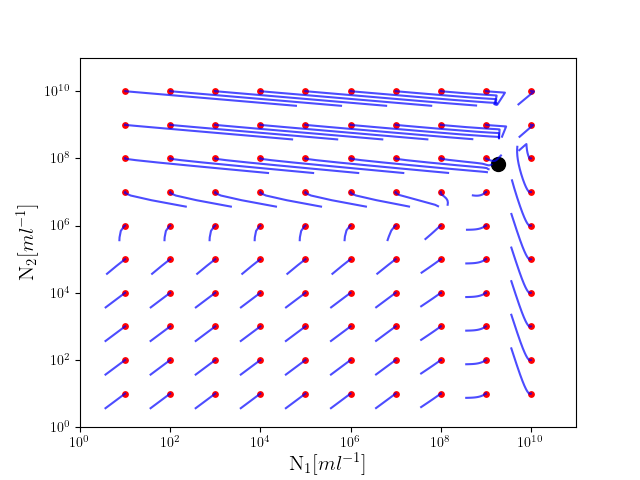}
	\caption{\textit{Phase space diagram of the case of $K_2 = 1 \times 10^{13}$. Note, how sufficiently high inoculation sizes are required for reaching the non-extinction state (black disc).}}
	\label{fig:phase_K1e13}
\end{figure}

Finally, we examine the influence of differences in metabolite affinity, expressed as the \textit{Monod} constant of the interacting populations (Table \ref{tab:K_2}). Again, the system shows a remarkable tolerance of differences between the two species . The phase space diagram of the most extreme case analysed ($K_2 = 1 \times 10^{13}$) again demonstrates the influence of the initial population sizes on the steady state the system converges to (Figure \ref{fig:phase_K1e13}). \\

\begin{table*}[!h]
	\begin{tabular} {l l l l l l}
		\hline \\[-0.7em]
		$K_2$ & $N_1(\infty)$ & $N_2(\infty)$ & $R(\infty)$ & $M_1(\infty)$ & $M_2(\infty)$ \\
		\hline \\[-0.6em]
		$1 \times 10^{1}$ & $ 6.4 \times 10^7$ & $ 1.9 \times 10^9$ & $ 5.6 \times 10^{9}$ & $ 5.8 \times 10^{3}$ & $5.8 \times 10^{12}$ \\		
		$1 \times 10^{5}$ & $ 6.4 \times 10^7$ & $ 1.9 \times 10^9$ & $ 5.6 \times 10^{9}$ & $ 5.8 \times 10^{7}$ & $5.8 \times 10^{12}$ \\
		$1 \times 10^{6}$ & $ 6.5 \times 10^7$ & $ 1.9 \times 10^9$ & $ 5.6 \times 10^{9}$ & $ 5.8 \times 10^{8}$ & $5.8 \times 10^{12}$ \\
		$1 \times 10^{7}$ & $ 6.6 \times 10^7$ & $ 1.9 \times 10^9$ & $ 5.6 \times 10^{9}$ & $ 5.8 \times 10^{9}$ & $5.8 \times 10^{12}$ \\
		$1 \times 10^{8}$ & $ 8.3 \times 10^7$ & $ 1.9 \times 10^9$ & $ 5.6 \times 10^{9}$ & $ 5.7 \times 10^{10}$ & $5.7 \times 10^{12}$ \\
		$1 \times 10^{9}$ & $ 2.3 \times 10^8$ & $ 1.8 \times 10^9$ & $ 5.6 \times 10^{9}$ & $ 5.3 \times 10^{11}$ & $5.3 \times 10^{12}$ \\
		$1 \times 10^{10}$ & $ 1.0 \times 10^9$ & $ 1.0 \times 10^9$ & $ 5.6 \times 10^{9}$ & $ 2.9 \times 10^{12}$ & $2.9 \times 10^{12}$ \\
		$1 \times 10^{11}$ & $ 1.8 \times 10^9$ & $ 2.3 \times 10^8$ & $ 5.7 \times 10^{9}$ & $ 5.3 \times 10^{12}$ & $5.3 \times 10^{11}$ \\
		$1 \times 10^{12}$ & $ 1.9 \times 10^9$ & $ 8.3 \times 10^7$ & $ 6.7 \times 10^{9}$ & $ 5.7 \times 10^{12}$ & $5.7 \times 10^{10}$ \\
		$1 \times 10^{13}$ & $ 1.9 \times 10^9$ & $ 6.6 \times 10^7$ & $ 1.9 \times 10^{10}$ & $ 5.7 \times 10^{12}$ & $5.7 \times 10^{9}$ \\
		$\geq 1 \times 10^{14}$ & \multicolumn{5}{l}{No non-trivial stable steady state} \\
		\hline  
	\end{tabular} \\[1.3em]
	\caption{\textit{The biologically feasible ($M_1(\infty), M2(\infty) > 0$) non-trivial stable steady state for varying metabolite affinities of the second species, expressed as their Monod constant $K_2$. All values are rounded to the first decimal place.}}
	\label{tab:K_2}
\end{table*}

\pagebreak

\subsection{Parameters relating to the limited resource}
Finally, we use the same approach to examine how differences in those biological traits, that relate to the utilisation of the limited resource $R$, will affect the equilibrium states of the system. From a biological point of view this is interesting because both species directly compete for this resource, whereas there is no competition for the exchanged metabolites. For this reason, differences in traits relating to the limited resource might affect the system in a different way than differences in traits which relate to the metabolites the symbionts provide each other with. For brevity purposes these results are presented in Appendix \ref{AppA}.

\section{The symbiotic spectrum}
\label{sec:spec}
In this section, we demonstrate how changes in the concentration of a metabolite can change the nature of the symbiotic interaction along a gradient ranging from true mutualism over commensalism to obligate parasitism. For this, we calculate the non-trivial stable equilibrium point of the system for different influx concentrations of the second metabolite using the standard parametrisation of the model described earlier. We compare the obtained population density equilibria with the equilibria obtained for the case of either of the two species being cultivated on its own. This way, we can assess whether the two populations gain an advantage or disadvantage from being cultivated together with the respective other one. We quantify these mutual effects by dividing the population density equilibria of the cocultivated populations by their respective counterparts in the 'monoculture' case. \\

In order to obtain the population densities for the 'monoculture' case, we re-use the devised general model after applying a number of simplifications. We here only show how to obtain the solution for $N_1$, since the case of $N_2$ is trivial. We first remove the second equation from equations (\ref{ode1_popsizes}), which describes the growth of the second population. The third term of equation (\ref{ode1_R}) is accordingly removed as well. Since we consider the case of $N_1$ growing alone, we are not interested in the concentration of the first metabolite, so we can remove the fourth equation of the model as well. Finally, the second term of the fifth equation can be removed as well, since $N_2$ equals zero. We are left with the following simplified model: \\

\begin{equation}
\begin{split}
\dot N_1(t) &= (\beta_1(M_2(t), R(t)) - \omega) \cdot N_1(t) \\
\dot R(t) &= \omega \cdot (R_{in} - R(t)) - \alpha_1 \cdot \beta_1(M_2(t), R(t)) \cdot N_1(t) \\
\dot M_2(t) &= \omega \cdot (M_{2, in} - M_2(t)) - \gamma_1 \cdot \beta_1(M_2(t), R(t)) \cdot N_1(t), \\
\end{split}
\end{equation}

which can be solved for $N_1(\infty)$ by solving the second equation for $R(\infty)$ and the third equation for $M_2(\infty)$ and substituting them in the first equation. Table \ref{tab:spec} summarises the results: Note, how an increasing influx concentration of the second metabolite diminishes the profit, which the first symbiont experiences from being cocultivated with the second species. At influx concentrations around $\approx 10 \times K_1$, the growth is not affected by the presence of the other symbiont, and for even higher concentrations the effect even becomes detrimental. In contrast, the second species is not able to survive on its own because the first metabolite is not supplemented into the culture. Therefore, it experiences a benefit from the presence of its symbiotic partner in all cases. Accordingly, by manipulating the external concentration of the second metabolite, we can shift the relationship between the two symbionts along a gradient which spans from obligate mutualism to obligate parasitism. \\

\begin{table*}[!h]
	\begin{tabular} {l l l l l l l l}
		\hline \\[-0.7em]
		 & \multicolumn{2}{c}{\underline {Cultivated alone}} & \multicolumn{2}{c}{\underline {Cocultivated}} & \multicolumn{2}{c}{\underline {Ratio}} \\
		$M_{2, in} \; [\times K_1]$ & $N_1(\infty)$ &  $N_2(\infty)$ & $N_1(\infty)$ & $N_2(\infty)$ & $N_1$ & $N_2$ & Interaction \\
		\hline \\[-0.6em]
		0 & 0 & 0 & $1.0 \times 10^9$ & $1.0 \times 10^9$ & $\infty$ & $\infty$ & Mutualistic \\
		0.1 & 0 & 0 & $1.0 \times 10^9$ & $1.0 \times 10^9$ & $\infty$ & $\infty$ & ---''--- \\
		0.2 & $8.6 \times 10^6$ & 0 & $1.0 \times 10^9$ & $1.0 \times 10^9$ & $116.3$ & $\infty$ & ---''---\\
		0.3 & $1.9 \times 10^7$ & 0 & $1.0 \times 10^9$ & $1.0 \times 10^9$ & $52.6$ & $\infty$ & ---''---\\
		0.4 & $2.9 \times 10^7$ & 0 & $1.0 \times 10^9$ & $1.0 \times 10^9$& $34.5$ & $\infty$ & ---''---\\
		0.5 & $3.9 \times 10^7$ & 0 & $1.0 \times 10^9$ & $1.0 \times 10^9$& $25.6$ & $\infty$ & ---''---\\
		1 & $8.9 \times 10^7$ & 0 & $1.0 \times 10^9$ & $1.0 \times 10^9$ & $11.2$ & $\infty$ & ---''---\\
		5 & $4.9 \times 10^8$ & 0 & $1.0 \times 10^9$ & $9.9 \times 10^8$ & $2.04$ & $\infty$ & ---''---\\
		10 & $9.9 \times 10^8$ & 0 & $1.0 \times 10^9$ & $9.8 \times 10^8$ & $0.99$ & $\infty$ & Commensalistic \\
		15 & $1.5 \times 10^9$ & 0 & $1.0 \times 10^9$ & $9.7 \times 10^8$ & $0.66$ & $\infty$ & Parasitic \\
		20 & $2.0 \times 10^9$ & 0 & $1.0 \times 10^9$ & $9.6 \times 10^8$ & $0.50 $ & $\infty$ & ---''--- \\
		\hline  
		
	\end{tabular} \\[1.3em]
	\caption{\textit{Population densities at the stable equilibrium for the two species when cultivated alone or together for different levels of metabolite influx concentration $M_{2, in}$. All non-zero population densities are rounded to the first decimal place.}}
	\label{tab:spec}
\end{table*}

The negative effect the first species expects as a consequence of the presence of the second one in case of high external concentrations of $M_2$ stems from the facts, that both species rely on the same limited resource for growth. Accordingly, changing the availability of this resource by manipulating its influx concentration $R_{in}$ can be expected to change the shape of the symbiotic spectrum. In order to analyse this in more detail, we will now compute the symbiotic spectrum for different influx concentration levels. Table \ref{tab:spec_Rin} summarises the results. Observe, how a higher resource availability shifts the point where the mutualistic relationship becomes a parasitic one towards higher influx concentrations of the metabolite. \\

\begin{table*}[!h]
	\begin{tabular} {l l l l}
		\hline \\[-0.7em]
		& \multicolumn{3}{c}{\underline {$\quad \qquad \qquad \qquad \qquad$ Ratio $N_1$ $\qquad \qquad \qquad \quad \quad$}} \\[0.25em]
		$M_{2, in} \; [\times K_1]$ & $R_{in} = 1 \times 10^{12}$ & $R_{in} = 2 \times 10^{12}$ & $R_{in} = 4 \times 10^{12}$ \\
		\hline \\[-0.6em]
		0 & $\infty$ & $\infty$ & $\infty$ \\
		0.1 & $\infty$ & $\infty$ & $\infty$\\
		0.2 & 58.1 & $116.3$ & 229.9 \\
		0.3 & 26.3 & $52.6$ & 105.3 \\
		0.4 & 17.2 & $34.5$ & 69.0\\
		0.5 & 12.8 & $25.6$ & 51.3\\
		1 & 5.62 & $11.2$ & 22.5\\
		5 & 1.04 & $2.04$ & 4.08\\
		10 & 0.52 & $0.99$ & 2.02\\
		15 & 0.35 & $0.66$ & 1.33\\
		20 & 0.27 & $0.50$ & 1.0\\
		\hline
	\end{tabular} \\[1.3em]
	\caption{\textit{The symbiotic spectrum for different levels of availability of the limited resource $R$. 'Ratio $N_1$' denotes the ratio between the stable population density of the first species when cocultivated together with its symbiont compared to a monoculture. Ratios above 1.0 indicate a mutualistic relationship, ratios below that level indicate parasitism.}}
\label{tab:spec_Rin}
\end{table*}

\section{Discussion}
\label{sec:disc}

In this work, a mechanistic model of disjunctive microbial symbioses based on metabolic interdependence has been derived. In contrast to preceding works, this model only relies on parameters which are directly measurable and understandable from biological first principles, instead of phenomenological terms such as 'cooperativity coefficients' that need to be fitted to experimental data \textit{a posteriori}. This way, the exact mechanisms underlying such symbiotic relationships are mechanistically understandable and concrete hypotheses can be generated which may be tested experimentally. Naturally, such models are less elegant from a purely mathematical point of view due to a higher number of parameters and equations, which makes them harder to solve analytically. Nonetheless, from a biological point of view they are able to provide valuable insights into the exact processes underpinning the interactions between the symbionts and thus make them a valuable tool for further research. \\

A main finding of this work is the remarkable stability of the modelled system towards differences in the biological traits of the involved organisms. This is a convenient finding from a perspective of applied microbiology (for example, see \citet{review_synthetic_communities}), because it shows that such systems can be constructed using organisms which might strongly differ from each other without endangering the long-term stability of the symbiotic relationship. This finding also might at least partially explain, why stable metabolic symbioses between microorganisms are not an uncommon sight in nature (for example refer to \citet{stable_associations, stable_associations_book, methanogenic_communities}). However, it needs to be noted that artificially establishing such a system \textit{in vitro} requires suitable starting conditions, $i.e.$ sufficient inoculation sizes of both organisms, or else the system will collapse and converge to the alternative stable steady state of permanent extinction. This also raises the possibility that higher inoculation sizes or the usage of smaller culture volumes may allow the cultivation of a number of microbial taxa previously thought unculturable, as also already noticed by \citet{GrowingUnculturableBacteria}. \\

In many cases, the production of complex substances comes with a diminished reproductive fitness \citep{metabolicburden}, as demonstrated by $e.g.$ the fitness costs bacteria experience as a consequence of antibiotic resistance mutations that rely on the activity of degradative enzymes \citep{fitnesscost_antibiotic}. Such effects are able to endanger the stability of the symbiotic association, if two conditions are met: First, the metabolite production and efflux rate must be subject to random mutation events, which is to be expected in biological systems. Second, genotypes with a higher metabolite efflux rate need to experience a monotonically increasing metabolic burden, $i.e.$ lower reproductive fitness. If both assumptions are met, it is easy to show that both populations will evolve towards ever decreasing metabolite efflux rates, if no effects are at play that are able to compensate for such differences in reproductive fitness. This would theoretically cause decreasing amounts of metabolites available for the respective symbiotic partner, increasingly slowing down microbial growth and ultimately the collapse of the system. Nonetheless, cases exist in which bacterial populations 'voluntarily' release metabolites in their environments and thereby uphold syntrophic relationships with other strains or species \citep{coli_split,methanogenic_communities, metabolic_interdependence}. The exact mechanisms behind this behaviour have not been understood yet and could therefore not be taken into account in the derivation of the presented model. Some proposed mechanisms include trade-offs in affinity for different resources, a trade-off between maximum growth rate and maximum biomass yield or a trade-off between resource uptake and biomass production \citep{syntrophy_definition}. Nonetheless, this problem remains far from solved. \\

The other interesting finding of this work lies in the demonstration of the existence of a continuous spectrum of symbiotic relationships, where mutualism and parasitism are only the most extreme cases of interaction. It could be shown that depending on the external metabolite availability, the exact same species interaction can fall on vastly different regions on this spectrum, which further substantiates the concept of symbioses as a continuous spectrum instead of static and isolated categories. This falls in line with both previous experimental \citep{syntrophy_spectrum, yeast_strains_metabolite_concentrations} and theoretical \citep{model_growth_rates, model_carrying_capacity} research. \\

In this work, only the case of disjunctive microbial symbiosis has been analysed. Throughout the scientific literature, several cases of conjunctive microbial symbioses based on metabolic interdependence have already been described \citep{jeon_I, jeon_II, stable_associations, stable_associations_book, m13_1, m13_2}. Additionally, such a physically associated microbial symbiosis was also artificially constructed in the laboratory and studied recently \citep{synthetic_stable_association}, however our knowledge of the exact growth and behaviour of such systems as a function of the environmental conditions and the biological traits of the involved symbionts is still too limited to model such systems reliably. 

\section{Acknowledgements}
I would like to thank my academic advisor Matthias Bild for helpful feedback during the derivation of the model. I am also indebted to Dr. C. v. T\"orne for his valuable support in the analytical examination of the model.

\bibliographystyle{apalike}
\bibliography{references}

\begin{thebibliography}{}

\bibitem[Canfield et~al., 2005]{stable_associations_book}
Canfield, D.~E., Kristensen, E., and Thamdrup, B. (2005).
\newblock {\em Aquatic geomicrobiology}.
\newblock Gulf Professional Publishing.

\bibitem[de~Bashan et~al., 2016]{synthetic_stable_association}
de~Bashan, L.~E., Mayali, X., Bebout, B.~M., Weber, P.~K., Detweiler, A.~M.,
  Hernandez, J.-P., Prufert-Bebout, L., and Bashan, Y. (2016).
\newblock Establishment of stable synthetic mutualism without co-evolution
  between microalgae and bacteria demonstrated by mutual transfer of
  metabolites (nanosims isotopic imaging) and persistent physical association
  (fluorescent in situ hybridization).
\newblock {\em Algal research}, 15:179--186.

\bibitem[Douglas, 1994]{def_symbioses}
Douglas, A. (1994).
\newblock {\em Symbiotic Interactions: Oxford Science Publications}.
\newblock Oxford University Press, Oxford.

\bibitem[Fagerbakke et~al., 1996]{ecoli_composition}
Fagerbakke, K.~M., Heldal, M., and Norland, S. (1996).
\newblock Content of carbon, nitrogen, oxygen, sulfur and phosphorus in native
  aquatic and cultured bacteria.
\newblock {\em Aquatic Microbial Ecology}, 10(1):15--27.

\bibitem[Graves et~al., 2006]{model_growth_rates}
Graves, W.~G., Peckham, B., and Pastor, J. (2006).
\newblock A bifurcation analysis of a differential equations model for
  mutualism.
\newblock {\em Bulletin of mathematical biology}, 68(8):1851--1872.

\bibitem[Gro{\ss}kopf and Soyer, 2014]{review_synthetic_communities}
Gro{\ss}kopf, T. and Soyer, O.~S. (2014).
\newblock Synthetic microbial communities.
\newblock {\em Current opinion in microbiology}, 18:72--77.

\bibitem[Hoek et~al., 2016]{yeast_strains_metabolite_concentrations}
Hoek, T.~A., Axelrod, K., Biancalani, T., Yurtsev, E.~A., Liu, J., and Gore, J.
  (2016).
\newblock Resource availability modulates the cooperative and competitive
  nature of a microbial cross-feeding mutualism.
\newblock {\em PLoS biology}, 14(8):e1002540.

\bibitem[Jeon and Jeon, 1976]{jeon_II}
Jeon, K. and Jeon, M. (1976).
\newblock Endosymbiosis in amoebae: recently established endosymbionts have
  become required cytoplasmic components.
\newblock {\em Journal of cellular physiology}, 89(2):337--344.

\bibitem[Jeon, 1972]{jeon_I}
Jeon, K.~W. (1972).
\newblock Development of cellular dependence on infective organisms:
  micrurgical studies in amoebas.
\newblock {\em Science}, 176(4039):1122--1123.

\bibitem[Kerner et~al., 2012]{tunable}
Kerner, A., Park, J., Williams, A., and Lin, X.~N. (2012).
\newblock A programmable {E}scherichia coli consortium via tunable symbiosis.
\newblock {\em PLoS One}, 7(3):e34032.

\bibitem[Kouzuma et~al., 2015]{methanogenic_communities}
Kouzuma, A., Kato, S., and Watanabe, K. (2015).
\newblock Microbial interspecies interactions: recent findings in syntrophic
  consortia.
\newblock {\em Frontiers in microbiology}, 6:477.

\bibitem[Madigan et~al., 2017]{brockmadigan}
Madigan, M.~T., Martinko, J.~M., Parker, J., et~al. (2017).
\newblock {\em Brock biology of microorganisms}, volume~13.
\newblock Pearson.

\bibitem[Mee et~al., 2014]{ecoli_strains_aa_syntrophy}
Mee, M.~T., Collins, J.~J., Church, G.~M., and Wang, H.~H. (2014).
\newblock Syntrophic exchange in synthetic microbial communities.
\newblock {\em Proceedings of the National Academy of Sciences},
  111(20):E2149--E2156.

\bibitem[Megee et~al., 1972]{syntrophy_spectrum}
Megee, R., Drake, J., Fredrickson, A., and Tsuchiya, H. (1972).
\newblock Studies in intermicrobial symbiosis. saccharomyces cerevisiae and
  lactobacillus casei.
\newblock {\em Canadian journal of microbiology}, 18(11):1733--1742.

\bibitem[Melnyk et~al., 2015]{fitnesscost_antibiotic}
Melnyk, A.~H., Wong, A., and Kassen, R. (2015).
\newblock The fitness costs of antibiotic resistance mutations.
\newblock {\em Evolutionary applications}, 8(3):273--283.

\bibitem[Miller, 1972]{m9}
Miller, J. (1972).
\newblock {\em Experiments in molecular genetics}.
\newblock Bacterial genetics - E. coli. Cold Spring Harbor Laboratory.

\bibitem[Monod, 1949]{monod}
Monod, J. (1949).
\newblock The growth of bacterial cultures.
\newblock {\em Annual Reviews in Microbiology}, 3(1):371--394.

\bibitem[Overmann and Schubert, 2002]{stable_associations}
Overmann, J. and Schubert, K. (2002).
\newblock Phototrophic consortia: model systems for symbiotic interrelations
  between prokaryotes.
\newblock {\em Archives of microbiology}, 177(3):201--208.

\bibitem[Pande and Kost, 2017]{metabolic_interdependence}
Pande, S. and Kost, C. (2017).
\newblock Bacterial unculturability and the formation of intercellular
  metabolic networks.
\newblock {\em Trends in microbiology}, 25(5):349--361.

\bibitem[Prescott et~al., 2010]{prescott}
Prescott, L.~M., Harley, J.~P., Klein, D.~A., and Willey, J.~M. (2010).
\newblock {\em Microbiologie}.
\newblock De Boeck Sup{\'e}rieur.

\bibitem[Raetz, 1996]{ecoli_dryweight}
Raetz, C. (1996).
\newblock Escherichia coli and salmonella cellular and molecular biology.
\newblock {\em Escherichia coli and Salmonella: Cellular and Molecular
  Biology}, 1:1035--1063.

\bibitem[Senn et~al., 1994]{ecoli_monod_for_glucose}
Senn, H., Lendenmann, U., Snozzi, M., Hamer, G., and Egli, T. (1994).
\newblock The growth of escherichia coli in glucose-limited chemostat cultures:
  a re-examination of the kinetics.
\newblock {\em Biochimica et Biophysica Acta (BBA)-General Subjects},
  1201(3):424--436.

\bibitem[Shapiro and Turner, 2018]{m13_2}
Shapiro, J.~W. and Turner, P.~E. (2018).
\newblock Evolution of mutualism from parasitism in experimental virus
  populations.
\newblock {\em Evolution}, 72(3):707--712.

\bibitem[Shapiro et~al., 2016]{m13_1}
Shapiro, J.~W., Williams, E.~S., and Turner, P.~E. (2016).
\newblock Evolution of parasitism and mutualism between filamentous phage m13
  and escherichia coli.
\newblock {\em PeerJ}, 4:e2060.

\bibitem[Shehata and Marr, 1971]{ecoli_phosphate_monod}
Shehata, T.~E. and Marr, A.~G. (1971).
\newblock Effect of nutrient concentration on the growth of escherichia coli.
\newblock {\em Journal of Bacteriology}, 107(1):210--216.

\bibitem[Shiloach and Fass, 2005]{ecoli_conversionrate}
Shiloach, J. and Fass, R. (2005).
\newblock Growing e. coli to high cell density—a historical perspective on
  method development.
\newblock {\em Biotechnology advances}, 23(5):345--357.

\bibitem[Stewart, 2012]{GrowingUnculturableBacteria}
Stewart, E.~J. (2012).
\newblock Growing unculturable bacteria.
\newblock {\em Journal of bacteriology}, 194(16):4151--4160.

\bibitem[Stump and Klausmeier, 2016]{syntrophy_definition}
Stump, S.~M. and Klausmeier, C.~A. (2016).
\newblock Competition and coexistence between a syntrophic consortium and a
  metabolic generalist, and its effect on productivity.
\newblock {\em Journal of theoretical biology}, 404:348--360.

\bibitem[Turner et~al., 1996]{coli_split}
Turner, P.~E., Souza, V., and Lenski, R.~E. (1996).
\newblock Tests of ecological mechanisms promoting the stable coexistence of
  two bacterial genotypes.
\newblock {\em Ecology}, 77(7):2119--2129.

\bibitem[Wu et~al., 2016]{metabolicburden}
Wu, G., Yan, Q., Jones, J.~A., Tang, Y.~J., Fong, S.~S., and Koffas, M.~A.
  (2016).
\newblock Metabolic burden: cornerstones in synthetic biology and metabolic
  engineering applications.
\newblock {\em Trends in biotechnology}, 34(8):652--664.

\bibitem[Yukalov et~al., 2012]{model_carrying_capacity}
Yukalov, V.~I., Yukalova, E., and Sornette, D. (2012).
\newblock Modeling symbiosis by interactions through species carrying
  capacities.
\newblock {\em Physica D: Nonlinear Phenomena}, 241(15):1270--1289.

\bibitem[Ziv et~al., 2013]{omega}
Ziv, N., Brandt, N.~J., and Gresham, D. (2013).
\newblock The use of chemostats in microbial systems biology.
\newblock {\em Journal of visualized experiments: JoVE}, (80).

\end{thebibliography}
\clearpage
\pagebreak

\appendix
\section{Further results}
\label{AppA}

We here present the additional results pertaining to the position of the equilibrium of the system, if the symbionts differ in those biological traits that relate to the utilisation of the limited resource $R$. Table \ref{tab:alpha_2} shows the influence of the resource requirement $\alpha_2$ of the second species. Increasing requirements will reduce the population density at the equilibrium point of the respective species, and, interestingly, will also reduce the population density of the other symbiont. This suggests, that changes in resource requirements will not give any advantage or disadvantage to either of the two symbionts, but will instead only change the overall carrying capacity of the system. This can also be seen from Equation (\ref{eq*}). \\

\begin{table*}[!h]
	\begin{tabular} {l l l l l l}
		\hline \\[-0.7em]
		$\alpha_2$ & $N_1(\infty)$ & $N_2(\infty)$ & $R(\infty)$ & $M_1(\infty)$ & $M_2(\infty)$ \\
		\hline \\[-0.6em]
		1 & $2.0 \times 10^9$ & $2.0 \times 10^9$ &  $5.6 \times 10^9$ & $5.8 \times 10^{12}$ & $5.8 \times 10^{12}$ \\
		$1 \times 10^1$ & $2.0 \times 10^9$ & $2.0 \times 10^9$ &  $5.6 \times 10^9$ & $5.7 \times 10^{12}$ & $5.7 \times 10^{12}$ \\
		$1 \times 10^2$ & $1.8 \times 10^9$ & $1.8 \times 10^9$ & $5.6 \times 10^9$ & $5.3 \times 10^{12}$ & $5.3 \times 10^{12}$ \\
		$1 \times 10^3$ & $ 1.0 \times 10^9$ & $ 1.0 \times 10^9$ & $ 5.6 \times 10^{9}$ & $ 2.9 \times 10^{12}$ & $2.9 \times 10^{12}$ \\
		$1 \times 10^4$ & $ 1.8 \times 10^8$ & $ 1.8 \times 10^8$ & $ 5.7 \times 10^{9}$ & $ 5.3 \times 10^{11}$ & $5.3 \times 10^{11}$ \\
		$1 \times 10^5$ & $ 2.7 \times 10^7$ & $ 2.7 \times 10^7$ & $ 6.7 \times 10^{9}$ & $ 5.7 \times 10^{10}$ & $5.7 \times 10^{10}$ \\
		\hline  		
	\end{tabular} \\[1.3em]
	\caption{\textit{The biologically feasible ($M_1(\infty), M2(\infty) > 0$) non-trivial stable steady state for varying resource requirements $\alpha_2$ per new cell of the second species. All values are rounded to the first decimal place.}}
	\label{tab:alpha_2}
\end{table*}

Table \ref{tab:L_2} shows the influence of the \textit{Monod} constant $L_2$ of the second species for the limited resource. Decreasing (increasing) the \textit{Monod} constant, $i.e.$ increasing (decreasing) the affinity of the species for the limited resource, will increase (decrease) its population size, however the effect on the symbiotic partner will vary. For small to intermediary values of $L_2$, an increase will benefit the symbiotic partner, which is likely caused by the higher levels of available resource not taken up by $N_2$. In contrast, in case of high values of $L_2$, the symbiotic partner $N_1$ will experience smaller population sizes as well, which is likely to be caused by a lack of produced and exported metabolite $M_2$. \\

\begin{table*}[!h]
	\begin{tabular} {l l l l l l}
		\hline \\[-0.7em]
		$L_2$ & $N_1(\infty)$ & $N_2(\infty)$ & $R(\infty)$ & $M_1(\infty)$ & $M_2(\infty)$ \\
		\hline \\[-0.6em]
		$1 \times 10^8$ & $ 6.5 \times 10^7$ & $ 1.9 \times 10^9$ & $ 5.6 \times 10^{9}$ & $ 1.1 \times 10^{9}$ & $5.8 \times 10^{12}$ \\
		$1 \times 10^9$ & $ 6.5 \times 10^7$ & $ 1.9 \times 10^9$ & $ 5.6 \times 10^{9}$ & $ 1.3 \times 10^{9}$ & $5.8 \times 10^{12}$ \\
		$1 \times 10^{10}$ & $ 6.6 \times 10^7$ & $ 1.9 \times 10^9$ & $ 5.6 \times 10^{9}$ & $ 3.9 \times 10^{9}$ & $5.8 \times 10^{12}$ \\
		$1 \times 10^{11}$ & $ 1.9 \times 10^9$ & $ 6.8 \times 10^7$ & $ 1.1 \times 10^{10}$ & $ 5.8 \times 10^{12}$ & $1.2 \times 10^{10}$ \\
		$1 \times 10^{12}$ & $ 1.8 \times 10^9$ & $ 6.1 \times 10^7$ & $ 1.1 \times 10^{11}$ & $ 5.5 \times 10^{12}$ & $1.7 \times 10^{9}$ \\
		$1 \times 10^{13}$ & $ 8.6 \times 10^8$ & $ 2.9 \times 10^7$ & $ 1.1 \times 10^{12}$ & $ 2.6 \times 10^{12}$ & $1.2 \times 10^{9}$ \\
		\hline  		
	\end{tabular} \\[1.3em]
	\caption{\textit{The biologically feasible ($M_1(\infty), M2(\infty) > 0$) non-trivial stable steady state for varying Monod constants $L_2$ of the second species. All values are rounded to the first decimal place.}}
	\label{tab:L_2}
\end{table*}

\end{document}